# Opportunistic Spectrum Sensing and Transmissions

*S. S. Alam, L. Marcenaro and C. Regazzoni*
Department of Biophysical and Electronic Engineering - DIBE
University of Genoa, Italy

**ABSTRACT**
Nowadays, cognitive radio is one of the most promising paradigms in the arena of wireless communications, as it aims at the proficient use of radio resources. Proper utilization of the radio spectrum requires dynamic spectrum accessing. To this end, spectrum sensing is undoubtedly necessary. In this chapter, various approaches for dynamic spectrum access scheme are presented, together with a survey of spectrum sensing methodologies for cognitive radio. Moreover, the challenges are analyzed that are associated with spectrum sensing and dynamic spectrum access techniques. Sensing beacon transmitted from different cognitive terminals creates significant interference to the primary users if proper precautions have not be not taken into consideration. Consequently, cognitive radio transmitter power control will be finally addressed to analyze energy efficiency aspects.

**Keywords:** *Cognitive radio networks, dynamic spectrum access (DSA), collaborative spectrum sensing, power control (PC), energy-efficient sensing.*

## 1. INTRODUCTION

The demand for wireless communications tremendously increased in time and to cope up with this demand, cognitive radio (CR) is a solution of huge prospect. CR can be described as an intelligent and dynamically reconfigurable radio that can adaptively regulate its internal parameters (or similar) in response to changes in the surrounding environment. CR has been made feasible by recent advances such as software-defined radio (SDR), machine learning techniques and smart antennas, etc (e.g., see Bixio et al, 2011). The use of CR technology allows in principle flexible and agile access to the spectrum as well as improving spectrum efficiency substantially. In this sense, CR represents a possible solution to the problem of spectrum scarcity, due to the variety of bandwidth demanding newly developed wireless communication techniques. In particular, spectrum scarcity occurs due to the use of traditional static frequency allocation planning for different communication protocols. Federal Communications Commission, FCC (2002) disclosed that the utilization of the continuously assigned spectrum still only ranges between 15% and 85%. This means that the primary users (also known as licensed or legacy users) do not often occupy the allocated radio resources (code, temporal and spatial domain) incessantly and this leads to their underutilization. Allocated but not used spectrum bands are usually known as spectrum holes or white spaces (WSs) in literature (Zhao & Sadler, 2007). WSs can opportunistically be brought into play in CR network by introducing efficient techniques expressed as dynamic spectrum access (DSA). DSA make it possible to a CR user to sense the vacant spectrum before using it temporarily, thus making it more likely to occur a improved efficient spectrum employment (Zhao & Sadler, 2007).

When a CR user aims at using WSs, interferences with the PU licensees may occur. Therefore, one of the main challenges in CR network is related to the management of the available radio resources among the PUs and cognitive users for satisfying the respective quality-of-service (QoS) requirements while limiting the interference to the PU licensees (Hsien-Po & Schaar, 2009). The growing interest of DSA in CR is specially related to the fact that it is considered as a possible solution of the static spectrum assignment. DSA techniques for the cognitive radio can be classified as *dynamic exclusive-use, shared-use or hierarchical access and open sharing or spectrum common models* as suggested by several researchers as E. Hossain et al (2009) and Couturier & Scheers (2009). In those papers DSA models were described; in the dynamic exclusive-use model, a licensed user can grant to a cognitive user the right to have exclusive access to the spectrum. In the hierarchical access model, an unlicensed user accesses the spectrum opportunistically without interrupting a licensed user when a legacy user is not interested to utilize that portion of the frequency spectrum. In an open sharing model, an unlicensed user can access the spectrum freely.

In order to dig up the benefit from dynamic spectrum access, knowledge about the spectrum WSs is necessary and this spectrum WSs can be discovered by *spectrum sensing,* one of the most important as well as challenging task, a CR has to perform. In particular, a CR should extract the information regarding the existence of active PUs in a specific frequency band and geographical location which is helpful and used to minimize the interference to licensed communications (Arshad et al, 2010). By sensing the spectrum, the terminal collects fundamental data from external environment, in particular through the sensing channel, and it can carry out the typical adaptation of CR (Gandetto & Regazzoni, 2007). The techniques of spectrum sensing in CR can generally be classified into stand alone or single transmitter based detection and distributed or cooperative fashion (Bixio et al, 2008). In their paper Bixio et al (2008) categorize different sensing techniques such as energy detection, matched filtering, radio identification, feature detection based and waveform based sensing as the different techniques which belong to transmitter based spectrum sensing; distributed detection with or without data-fusion (Cattoni et al, 2006) are addressed as sub-classes of feature based sensing techniques when cooperative spectrum sensing is performed. In CR environment cooperative sensing is a potential solution of the type hidden terminal problem is similar to other WLAN (IEEE 802.11) schemes. In spite of the enormous advances made on stand-alone techniques, spectrum sensing remains a complex task in real cognitive radio environment due to the multipath fading, frequency selectivity, time varying channels and noise (Yücek & Arslan, 2010).

DSA techniques, despite enhancing overall spectrum utilization, can carry to the decision of radio signal transmission by cognitive nodes that can produce harmful interference to the PUs. This aspect must be taken into account: to this end, two important design criteria for CR networks have been fixed, i.e., joint maximization of radio resource utilization and minimization of the interference caused to the primary users (Hoang & Liang, 2008). Consequently, power control techniques assumed great importance in CR communication to deal with the minimization of the probable cross-interference originated by multiple users, either primary or secondary, within CR networks. Generally, power control is employed in mobile networks for improving link performance (e.g., see DS-CDMA systems); similar concepts can be applied to CR networks as well. Power control is a crucial task also in CR networks and it is one of the basic modules of radio resource management (Qin & Su, 2009). In a wireless network where users are prioritized (i.e. both PUs and CRs are present), if a CR gets the access to exploit the spectrum, it must take into account the power transmission itself to assure an interference-free transmission for the surrounding users (PUs and CRs both). There are different techniques of this kind. For example, spectrum underlay is one of the existing techniques which permits cognitive users to share the whole spectrum with the PUs as long as they do not exceed the interference threshold at PUs (Pareek et al, 2010). Shi and Hou (2007) have been demonstrated an excellent survey on power control of cognitive radio and this is followed by the known "*protocol model*" for interference modeling. Since power control directly affects the transmission range and interference range (received power at the destination node and interference power at other nodes) of a cognitive terminal, it has profound impact on scheduling feasibility, bandwidth efficiency, and problem complexity (Shi & Hou, 2008). Furthermore, power control is an important issue for the energy-efficient spectrum sensing, the future trend of green cognitive radios.

The remainder of this chapter is organized as follows: section two is engaged with the problems related to the traditional static spectrum allocation, in section three, a comprehensive overview of the DSA techniques in CR networks, focus will be on the state of the art in DSA, pros and cons of each DSA sub-classes are drawn, then spectrum sensing is discussed in section four and then optimal, energy efficient power allocation, which will be helpful to mitigate the interference temperature, will be discussed. At last, some conclusions will be drawn.

## 2. SPECTRUM SCARCITY PROBLEMS AND WHY DSA COMES OUT

Nowadays, the availability of frequency spectrums are running out due to the use of traditional static spectrum allocation (SSA) strategy and a growing number of wireless applications and equipments. The frequency spectrum is assigned worldwide by the world radio communications conference (WRC) under the presence of all the member nations as they demand the frequency and of different service operators that require spectrum allocations for mobile, fixed, and satellite technologies. The government authority of each country is responsible to assign the portion of spectrums for different communication vendors, operators or organizations for different purposes. The long-term spectrum allocation is granted in space and time invariant and any changes to it happen under strict control of the regulatory authority. The licensee is responsible for the fixed spectrum whether the spectrum is used or not.

The SSA scheme assigns a specific spectrum for a communication system to be employed for a certain geographic location. This allocation is fixed and cannot be changed over time and space. A large part of the radio spectrum is allocated but barely used in most of the locations and time. The portions of the spectrum which are temporarily unused by the licensee, are called the spectrum holes or spectrum white spaces (WSs) or vacant spaces and that WSs present in temporal, frequency and spatial domains. Several radio bands allocated for military, government and public safety use experience negligible utilization. FCC (2002) disclosed the static partitioning of spectrum has significant operational implications which have been recently brought to light by extensive spectrum utilization measurements in the USA and Europe. The available spectrum can be divided into licensed and license-exempt frequency bands. Usually, licensed frequency spectrum provides for some degree of interference protection which is called guard band because each new licensee must demonstrate compliance with certain standards for limiting interference to other existing nearby licensed systems. The radio parameters are defined as those parameters (e.g., radiated transmitter power level, frequency, SNR, modulation techniques, etc) that may have influence to each licensee. License-exempt bands do not require individual transmitters to be licensed in order to operate, but there are still radiated power restrictions that usually keep power at low levels as a de facto way of limiting interference.

SSA has several shortcomings because of being time and space invariant. In SSA, parts of the radio spectrum have been statically allocated for various wireless networking services for the military, government, commercial, private and public safety systems. As noticed by FCC (2002), the main problem of static spectrum allocation is the underutilization of the radio spectrum, as revealed by extensive measurements of actual spectrum usage; this fact caused several activities in the engineering, economics, and regulation communities in searching for better spectrum management policies. Often, the spectrum reserved by the government authority for a particular purpose could be not in use. Another major problem with the static spectrum allocation and the legacy static radio technology is the lack of interoperability with existing technologies. Interoperability is hampered by the use of multiple frequency bands, incompatible radio equipment, and a lack of standardization. Therefore, first responders (or emergency responders i.e., police, fire or emergency medical services) from different jurisdictions and agencies often cannot communicate during emergencies (Wang et al, 2011). There have been examples of failure in communication between different organizations at world trade center (WTC) on 9/11/2001, which is mentioned in several open literatures. For example, some of the police warnings regarding to the evacuation of people were not received in the area of the second building of WTC. Unfortunately, the fire fighters did not receive alarm messages thus leading to considerable loss of life and other disaster areas due to lack of interoperability among the primary/legacy radios (Dilmaghani & Rao, 2006). The other problem of static frequency allocation is that it cannot change the assigned spectrum properly even if the transmission channel is noisy although there are other options to choose the available spectrum (for less noisy transmission) which is not being used at that instant. Generally, the spectrum usage pattern for military intentions and commercial systems are quite different. It may certainly happen that the usage of spectrum in certain networks is lower than anticipated, while in other location users survive for the spectrum because of high demand in that geographic location. Static allocation of frequency spectrum fails to deal with this kind of issues of spectrum sharing. Static spectrum allocation additionally faces difficulties due to the modification of old technologies. In order to overcome the problems related to SSA, one should move towards the regime of DSA, where flexibility, efficient utilization of spectrum is concerned. DSA and its different possibilities are elaborately expressed in the following sections.

### 3. DYNAMIC SPECTRUM ACCESS (DSA) IN CR NETWORKS

In order to meet the massive demand of frequency spectrum, the CR network has opened up a new way of sensing and utilizing properly wireless radio resources (spectrum both in temporal and spatial domain). Cognitive radio is a dynamically reconfigurable radio which can adjust its radio parameters in response to the surrounding environment. This has been made feasible by recent advances such as software-defined radio (SDR) and smart antennas (Bixio et al, 2010). By using such CR devices enables flexible and agile access to the wireless spectrum, which can in turn, improve efficiency in spectrum utilization significantly. In this section, the state of art DSA schemes will be discussed and many researches are summarized that have been carried out using those techniques in the context of distributed CR networks. Nowadays, wireless communication is suffered from spectrum scarcity due to newly developed communication techniques always required an additional exclusive spectrum access. Latest experiment on spectrum management (Akyildiz et al, 2006) have shown that the licensed frequency bands are rigorously underutilized most of the time and a particular geographic location mainly due to the traditional command and control type spectrum regulation (i.e., static spectrum regulation) that has prevailed for decades. In order to use those remaining spectrum holes or white spaces, effort is put on achieving DSA.

Under such a spectrum policy, each spectrum band is assigned to a designated party, which is given an exclusive spectrum usage right for a specific type of service and radio device. CR can manage in order to mitigate the spectrum scarcity problem by enabling DSA scheme, which allows CRs to identify the unemployed portions of licensed band and utilize them opportunistically as long as the CRs do not interfere with the PUs communication. The diversity of the envisioned spectrum reform ideas is manifested in the number of technical terms coined so far: dynamic spectrum access vs. dynamic spectrum allocation, spectrum property rights vs. spectrum commons, opportunistic spectrum access vs. spectrum pooling, spectrum underlay vs. spectrum overlay. A taxonomy of the DSA scheme (Zhao & Swami, 2007) is illustrated in the following figure (fig. 1).

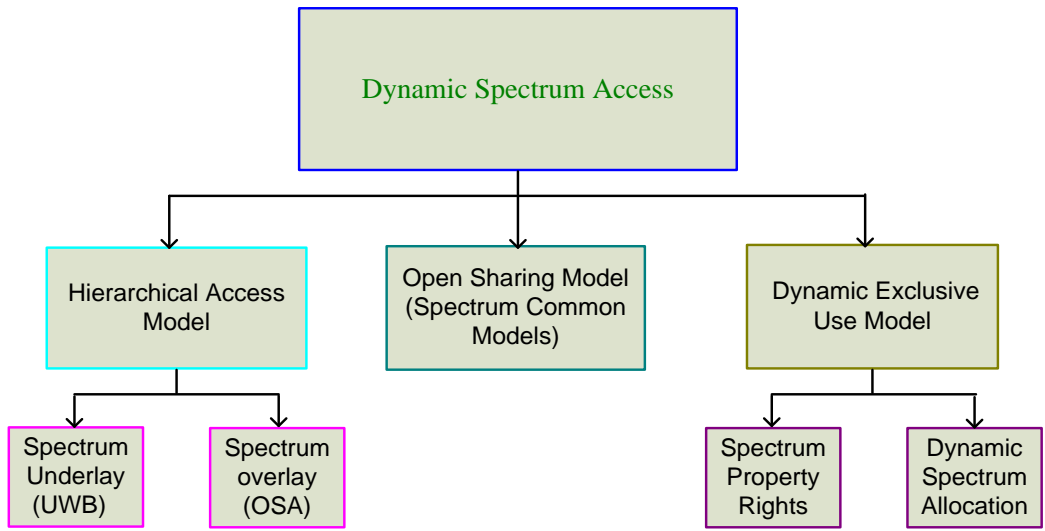

Fig.1: Fundamental classification of dynamic spectrum access

### 3.1 Hierarchical Access Model

In this model, a hierarchical access pattern for the primary and secondary users will be discussed. The fundamental concept is to open licensed spectrum to cognitive users while limiting the interference perceived by the primary users. This model can be categorized as two different approaches for allocation of the spectrum, i.e., spectrum underlay and spectrum overlay. Both the schemes motivate with the matching goal that CR users need to have an overview about the current spectrum utilization in order to detect and identify PUs. Spectrum underlay exploits the spectrum by using it despite of a PU transmission, but by controlling the interference within a prescribed limits. This can be obtained by using spread spectrum techniques, resulting in a signal with large bandwidth but low spectral power density, which can coexist with primary users. Spectrum overlay intends to use spectrum holes in an opportunistic way (e.g., opportunistic spectrum access, OSA) while without interfering PUs transmission, indicating that the spectrum is periodically monitored by the secondary or cognitive user seeking absence of PUs in order to utilize the vacant spectrum by the CR user.

#### 3.1.1 Spectrum Underlay

In an underlay system, regulated spectral masks impose stringent limits on radiated power as a function of frequency, and perhaps location. Radios coexist in the same band with primary licensees, but are regulated to cause interference below prescribed limits (Zhao & Swami, 2007). For simplicity, consider a low-powered radio that could coexist within the same frequency channel with a high-powered broadcast radio. Because of the power limitation, underlay radios (URs) must spread their signals across large bandwidths spreading with lower energy, and/or operate at relatively low rates. UWB radio could be an example of this kind. The power limitation results in a corresponding limit on rate-range capabilities. An advantage of such a system is that radios can be dumb, they do not need to sense the channel in order to defer to primary users. The underlying principle is that the primary users are either sufficiently narrow-band, or sufficiently high-powered, or the URs are sufficiently fast frequency hopping with relatively narrow bandwidth usage in each dwell, so that there is little interference from the URs. An underlay radio spectrum distribution is shown in the following figure (fig. 2).

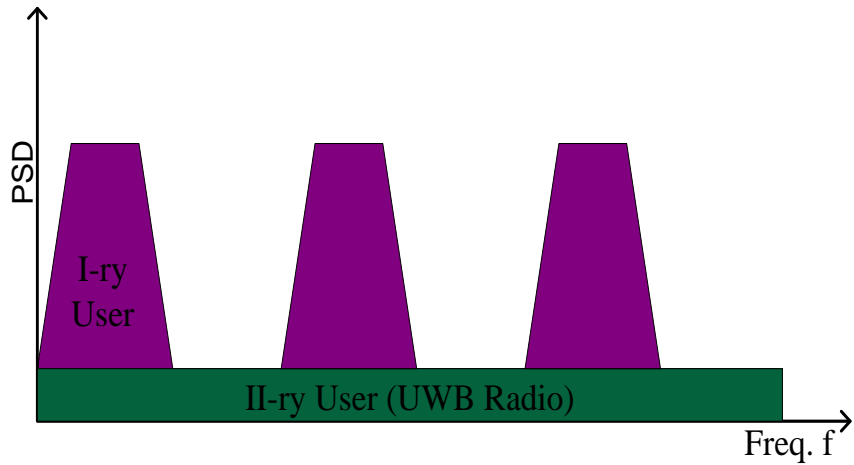

Fig. 2: Spectrum Underlay

In order to spread out the signal over a large bandwidth, underlay radios can use spread spectrum signalling systems, wideband orthogonal frequency division multiplexing (OFDM), or impulse radio. Because of the large front-end bandwidth, URs are susceptible to interference from a sort of co-existing sources, including relatively narrowband signals from primary users. This can cause saturation of the automatic gain control (AGC) circuit leading to signal distortion and loss of dynamic range. It is important to notice that suppressing strong primary signals through front-end notch filters is complex, since there could be many primary signals, and not always at the same frequency locations. Receiver arrays can help notch some primary users by exploiting the spatial degrees of freedom. A second problem is that high-resolution high-rate analog to digital converter (ADC) is extremely challenging due to both the high power consumption of such devices and fundamental limits imposed by the noise level. Consequently, it may be necessary to devise and implement analog or digital converters to achieve high-fidelity sampling at a rate slower than the system bandwidth. Current URs have limited range and rate, not so convenient to deal with the aggregate interference and have largely been confined to indoor applications. Floor URs must also be capable of dealing with the large delay spread and frequency selectivity of the channel. In short, URs tend to be complex in terms of hardware implementation, but relatively dumb in terms of spectrum sensing and access protocols. Challenges exist in hardware implementation, frontend interference suppression, high-fidelity low-power high-rate ADC circuit design, and estimation and equalization of long delay-spread channels.

Modelling of aggregate interference from URs and devising algorithms to cope up with the primary receivers have not been adequately addressed. Another aspect of aggregate interference is that spectral masks may have to be adapted to secondary traffic load. An UR could sense the spectrum so as to shape its transmission signal to avoid congested bands. This requires reliable sensing of the spectrum similar to the spectrum overlay systems discussed in the following section.

### 3.1.2 Spectrum Overlay

Spectrum overlay or OSA, can be applied in either temporal or spatial domain. For the first case, secondary users aim to exploit temporal spectrum opportunities resulting from the bursty traffic of primary users and in the latter, cognitive users aim to exploit frequency bands that are not used by primary users in a particular geographic area (Zhao & Swami, 2007). A typical application is the reuse of certain TV white spaces that are not used for TV broadcasting (e.g., digital TV transmission) in a particular geographic location. In the TV broadcasting system, TV-bands assigned to adjacent regions are different to avoid co-site interference. This results in unused frequency bands varying over space. In general, spectrum opportunities vary in both temporal and spatial domains. It is often assumed in the literature that one variation is at a much slower scale than the other. Spectrum overlay mechanism is shown in the following (fig. 3).

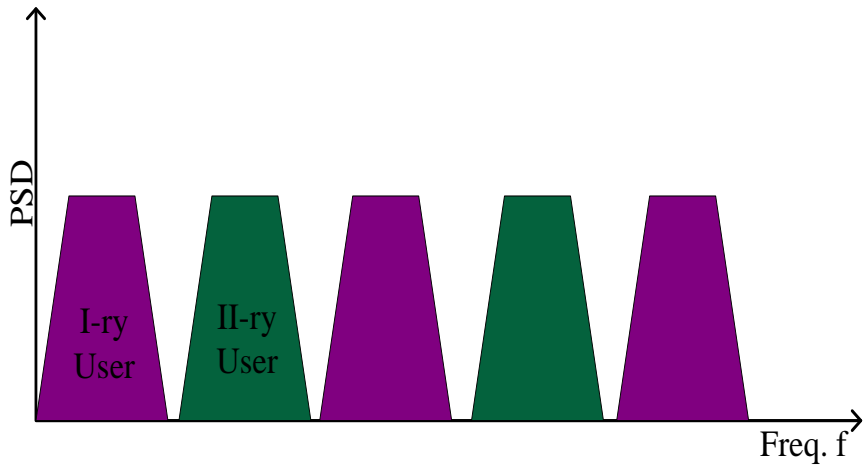

Fig. 3: Spectrum Overlay (e.g. Spectrum Pooling or OSA)

OSA is also known as interweaving of frequencies in some papers. Spectrum overlay is therefore defined as doing some pre-coding at the transmitter in order to diminish the interference at the receiver. Therefore extensive knowledge about other signals in the specific spectrum is necessary. This technique is also known as ***dirty paper coding*** (Erez & Brink, 2004). The majority of existing work on OSA focuses on the spatial domain where spectrum opportunities are considered static or slowly varying in time. As a consequence, real-time opportunity identification is not as critical a component in this class of applications, and the prevailing approach tackles network design in two separate steps: (i) opportunity identification assuming continuous full spectrum sensing; (ii) opportunity allocation among secondary users assuming perfect knowledge of spectrum opportunities at any location over the entire spectrum.

To find out the spectrum opportunity in the presence of fading and noise there will be some uncertainty and this issue is discussed in some papers and spatial opportunistic allocation is also possible among secondary users. OSA in the time domain need to design both of the spectrum sensing and access schemes. Tracking the rapidly varying spectrum opportunities would be a crucial issue, and a simple yet sufficiently accurate statistical model of spectrum occupancy is crucial to the efficiency of spectrum opportunity tracking. There are some errors predictable in real-time sensing, and the characteristics of the spectrum sensor should be taken into account in making spectrum access decisions. Initial attempts at addressing the identification and exploitation of temporal spectrum opportunities that also vary in space and for an extended overview of challenges and recent developments in OSA, can be investigated by Zhao & Sadler (2007) in their research paper.

Basically, a channel can be considered as an opportunity if it is not currently being used by legacy/primary users. As suggested by the Zhao & Sadler (2007), the spectrum opportunity can be defined as follows.

a) Spectrum opportunity is a local concept expressed as to a particular pair of cognitive users depending not only on the location of the cognitive transmitters but also on the position of secondary receivers. For multicast and broadcast, spectrum opportunity is open for interpretation, and results in networking tradeoffs.
b) Spectrum opportunity is determined by the communication activities of PUs rather than that of cognitive users. Failed communications originated by collisions among SUs do not disqualify a channel from being an opportunity.

## 3.2 Dynamic Exclusive Use Model

In the exclusive-use model for spectrum access, the radio spectrum is licensed to a user/service to be exclusively used under an agreement. This model maintains the basic structure of the current spectrum regulation policy: spectrum bands are licensed to services for exclusive use. The main idea is to introduce flexibility to improve spectrum efficiency. Two schemes have been proposed under this model: spectrum property rights and dynamic spectrum allocation (Couturier & Scheers, 2009).

Generally, PUs do not utilize their spectrum all the time even if they deserve this spectrum to use. Therefore, the PUs can sub-lease those underutilized spectrum to third party thus leading to the spectrum trading. This type of spectrum trading can be given the right to exclusively use those resources without being mandated by a regulation authority. This approach is called Spectrum Property Rights, as the license or the right is based on the three spectrum properties: fixed frequency band, time and a geographic location.

### 3.2.1 Spectrum Property Rights

The detailed explanation of spectrum property rights and challenges of this scheme can be found in (Hatfield & Weiser, 2005). Specifically, L. White (2000) stated in his paper that "*the property right would be expressed as the right to transmit over the specified spectrum band, so long as the signals do not exceed a specified strength beyond the specified geographic boundaries during the specified time period*". One of the major difficulties in enforcing such spectrum property rights lies in the unpredictability of radio wave propagation in both frequency and space. Spectral and spatial spillover is inevitable, unpredictable, and depending on the characteristics of both transmitters (potential trespassers) and receivers (property right owners).

### 3.2.2 Dynamic Spectrum Allocation

A second approach for the dynamic exclusive use model is dynamic spectrum allocation. For this the temporal and spatial traffic statistics are exploited, which is valuable for sub-leasing long-term applications, such as UMTS or DVB-T. Sub-leasing based on traffic statistics leads to a much more flexible spectrum allocation than in the previous approach. For example, the amount of spectrum allocated to UMTS and DVB-T can vary over region and the-time-of-day. But again, dynamic is limited to the capabilities of the licensee, so it is unlikely that with either of these approaches the spectrum holes can be optimally filled (Couturier & Scheers, 2009).

The literature of dynamic spectrum allocation is extensively discussed by L. Xu et al (2000). This work is related to European Union funded project DRiVE (**D**ynamic **R**adio for **I**P Services in **V**ehicular **E**nvironments), aiming at enabling spectrum-efficient high-quality wireless IP communication in a heterogeneous multi-radio networks considering a common co-ordination channel. DSA opens new possibilities for wireless network operators if a multi-radio infrastructure for optimized interworking of cellular and broadcast network is considered:

- Operators can allocate spectrum for each service access network according to local and temporal needs.
- Users on the move are provided with the benefit of accessing enhanced IP based mobile services on the fly and wherever they are in a cost efficient way.

Multiple networks regulation policy and issues in the context of dynamic spectrum allocation are pointed out in (Leaves et al, 2004). Temporal and Spatial DSA algorithms are discussed there and several implementations of an iterative, real-time, temporal DSA algorithm are possible, but typical steps in the operation of the temporal DSA algorithm include: a) Periodic triggering of DSA algorithm, b) Management of the traffic on the carriers, c) Prediction of the loads on the networks and d) Allocation decision while the goal of spatial DSA is to allocate spectrum to radio access networks (RANs) according to the traffic requirements in each location using DSA scheme.

Furthermore, it is necessary to coordinate the spectrum allocation between adjacent DSA areas for avoiding interferences. Especially, the spectrum allocations of different RANs belonging to adjacent DSA areas should not overlap in the same portion of spectrum. In order to avoid this spectrum overlap while still allowing spectrum allocation adaptation to the traffic demand, the guard band needs to be increased to guarantee the coexistence of the different systems. The structure of an usual spatial DSA scheme can be summarized in three main steps: *a) calculating the spectrum overlap, b) performing initial assignment and c) optimize the spectrum usage*.

Two centralized dynamic spectrum allocation (Recursive Round Robin Algorithm and Greedy-Type Algorithm) protocols that rely on a super base-station and their performance evaluated via simulations are depicted by Aazhang et al (2004). Two simple protocols to redistribute call loading among base stations and facilitate sharing of the spectrum is analyzed in that paper. The protocols proved to be simple yet effective. The protocols also exhibited a tradeoff among fairness, complexity, effectiveness and possibly introducing call handling delays in the cellular network.

### 3.3 Open Sharing Model

The dynamic exclusive use model and the hierarchical access model concern with the primary users having a license to use a certain part of the spectrum, whereas the open sharing model assumes a vacant spectrum with only peer users. Compared to the other two models, many technical issues under this model are perhaps the closest to the conventional medium access control (MAC) problems. Again, two different approaches how to organize interference-free communication are illustrated by Couturier and Scheers (2009). This open sharing model can be categorized as centralized and distributed fashions. Centralized and distributed spectrum sharing strategies have been considered to address the technological challenges under this model.

In a *centralized open sharing model*, there is only one centralized cognitive manager that controls the whole cognitive radio domain. The cognitive manager can be straightforwardly implemented using an expert system, or the problem can be seen as an optimization problem for which a global optimum has to be found. The centralized approach assumes however that there is a reliable cognitive signalling channel connecting each radio to the centralized manager. It has to be noted that the cognitive manager, in general, will not only influence the proficient spectrum usage, but also other transmission parameters like transmit-power, signal to noise ratio, modulation strategy, etc.

In the distributed open sharing model, decision making is more complicated. Decisions have to be taken locally by all the transmitter-receiver pair, meaning that there must be a cognitive manager in every node. In this case, coordination between pairs or coalitions of pairs can facilitate the spectrum sensing, competent use of radio resources and enhance the quality of the information, on which the pairs can rely to make their decisions.

#### 3.3.1 Centralized DSA

In a centralized dynamic spectrum access architecture, a central controller is deployed to collect and process information about the radio environment. With a central controller, the decision of cognitive radio users to access the spectrum can be made such that the desired system-wide objectives are achieved. In a centralized scheme, every CR user always communicates with a central controller to inform their status and in response to that the central controller update to every CR node. The central controller is inside CR network and makes the decisions intelligently as the requirement of CR user. There are two approaches proposed to implement centralized dynamic spectrum access namely, optimization approach and auction-based approach (Hossain et al, p. 274, 2009). With an optimization-based approach, different types of optimization problems can be formulated (e.g., convex optimization, assignment problem, linear programming, and graph theory). Standard methods in optimization theory can then be applied to obtain the optimal solution for dynamic spectrum access. While auction theory driven centralized dynamic spectrum access mainly address the spectrum trading in a suitable way. In this approach, CR users submit their bids to the spectrum owner for each vacant spectrum and the highest bidder will then win to utilize the spectrum accordingly.

### 3.3.2 Distributed DSA

In many scenarios such as in ad-hoc cognitive radio networks, deploying a central controller may not be feasible (Hossain et al, 2009). Therefore, distributed dynamic spectrum access would be required in such cognitive radio networks. Due to the absence of any central controller, each unlicensed user has to gather, exchange, and process the information about the wireless environment independently. Also, an unlicensed user has to make decisions autonomously based on the information that available on the radio environment to access the spectrum, so that the unlicensed user can achieve its performance objective under interference constraints. The common behaviors of an unlicensed user in a cognitive radio network without a central controller are as cooperative or non-cooperative behavior, collaborative or non-collaborative behavior and learning ability.

#### 3.3.2.1 Cooperative or non-cooperative behavior

Since a central controller which controls a decision of spectrum sharing is not available, a CR user can adopt either cooperative or non-cooperative behavior. A CR with cooperative behavior will make a decision on spectrum access concerning the performance of the overall network (i.e., a group objective), though this decision may not result in the highest individual benefit for each CR user. In contrast, a CR user with non-cooperative behavior will make a decision that is opposite to cooperative behavior i.e. it wants to maximize the individual performance whatever would be the network performance. This feature of the CR user is also called the selfish behavior. Couturier and Scheers (2009) discussed in their paper game theory and iterative water filling approach can be used for the distributed DSA.

To pertain game theory to the process of decision making in a cognitive radio, the decision making process needs to be modeled as a game. First of all, it should be checked whether it is a centralized or a distributed DSA model (i.e., the centralized or the distributed open sharing model). Secondly, it must be decided which performance metric(i.e., the throughput or the latency) is to be optimized. Thirdly, all information about any cognitive radio in the environment of the decision maker needs to be collected (i.e., the possible actions and the preferred strategy).

Finally, a mapping of the elements of a cognitive radio to a game must be carried out, as depicted in *table-1* which shows the number of players can be mapped to the number of transmitter-receiver groups, where a group consists of one transmitter and an arbitrary number of receivers (Couturier & Scheers, 2009).

| Game Theory | Variable | Cognitive Radio |
|---|---|---|
| No. of Players | M | No. of nodes to be considered |
| Entirety of actions | A | Transmission parameters sets |
| Utility function, Payoff | U | Performance metrics |

**Table 1**: Mapping of cognitive radio elements to plot a game (Couturier & Scheers, 2009)

In general, the CR users are in the same group with the same objective, these users typically choose cooperation among them but if the CRs be independent and have different objectives, non-cooperative behavior would much likely be practiced.

### 3.3.2.2 Collaborative or Non-collaborative Behavior

Again, since a central controller which can coordinate the gathering and broadcasting of network information is not available, the CR user has to collect network information themselves. In this case, a CR can be collaborative or non-collaborative in nature to exchange network information for distributed dynamic spectrum access. With non-collaborative behavior, all network information is gathered and processed locally by each CR user when there is no interaction among the secondary users. Conversely, with collaborative behavior, the CR user can exchange network information with each other. Again, the choice to become either collaborative or non-collaborative depends on the type of the cognitive radio network and the users. Also, the collaboration of the SUs relates to cooperative and non-cooperative behavior. Typically, collaboration among CR users to exchange network information is required to achieve cooperative behavior. Also, if the CRs are non-collaborative, they are also typically non-cooperative. However, if the CRs are collaborative, they could be either cooperative or non-cooperative. For example, the unlicensed users may agree to reveal some information (e.g., the chosen spectrum access action), but they make a decision to achieve their own objectives (i.e., non-cooperative), rather than a group objective. Therefore, an unlicensed user can be classified into one of the following three categories: *collaborative–cooperative*, *collaborative–non-cooperative,* and *non-collaborative–non-cooperative*. In the case of collaboration among unlicensed users, a protocol will be required for exchanging network information. However, in the case of non-collaborative unlicensed users, the network information has to be observed and learned locally. Therefore, learning ability will be crucial for a cognitive radio which is discussed in the following.

### 3.3.2.3 Learning Ability

The ability to learn and make intelligent decisions is important for distributed dynamic spectrum access. In particular, the CRs have to observe and learn the system state (e.g., the occupancy of the radio spectrum in a certain time). The output of this learning process is the knowledge about the radio environment and the system, which would be useful for a CR user to make a decision on spectrum access. Again, the learning process can be either non-collaborative or collaborative. In the case of non-collaborative learning, the knowledge about the system is produced by each individual unlicensed user without interaction with other users. On the other hand, the unlicensed users can collaborate not only to exchange network information, but also to process and produce the system knowledge. Then, based on this knowledge, an unlicensed user can make the decision whether to achieve the group objective or its individual objective.

### 3.4 Medium Access Control (MAC) for DSA

Basic components of cognitive MAC basic design components of cognitive MAC for the opportunistic spectrum access include a sensing policy for real-time decisions which determines the proper sensing instant and an access guidelines that determines how to access the spectrum based on the sensing outcomes (Hossain & Bhargava, p. 272, 2007). The purpose of the sensing policy focus on two aspects: to identify a spectrum opportunity for immediate access and to obtain statistical information on spectrum occupancy for improved future decisions. A trade-off should be considered between these two often conflicting objectives, and the trade-off should adapt to the bursty traffic and energy constraint of the CR user. For example, when there are energy costs associated with sensing, a secondary user may decide to skip sensing when its current estimate of spectrum occupancy indicates that no channels are likely to be idle. Clearly, such decisions should balance the reward in energy savings with the cost in lost spectrum information and potentially missed spectrum opportunities. The objective of the access policy, on the other hand, is to minimize the chance of overlooking an opportunity without violating the constraint of being non-intrusive. Whether the secondary user should adopt an aggressive or a conservative access policy depends on the operating characteristics (probability of false alarm vs. probability of miss detection, and tolerable level of interference) of the spectrum sensor. A joint design of MAC protocols and spectrum sensors at the physical layer is thus necessary to achieve optimality. Energy constraints will further complicate the design of access policies. For energy-constrained OSA in fading environments, the secondary user may avoid transmission when the sensed channel is in a deep fade. Even the residual energy level will play an important role in decision-making. A detailed survey of the essential MAC features for the opportunistic spectrum access is carried out by Pawełczak et al (2008).

### 3.5 Open issues in DSA

The fundamental concepts of DSA scheme in a cognitive radio network have been discussed in the above section. Different spectrum access models for cognitive radio, namely, the exclusive use, shared-use, and commons models, have been described. Spectrum sensing and spectrum access are the two major functionalities of a cognitive MAC protocol so a brief study on the cognitive MAC protocols has been expressed. DSA poses different sets of challenges in centralized and distributed cognitive radio networks. There are lot of works on sensing have performed even though sensing itself can cause interference to the PUs so the sensing with the learning algorithms are used to build knowledge about both the ambient radio environment and the network based on the observations. This knowledge is then used by the cognitive radio users to make spectrum access decisions. In addition, to support distributed DSA, various signaling protocols are used for exchanging network information between the collaborative unlicensed users.

Above all, there are many issues that have not been standardized yet. The problems associated with the scheduling problems, QoS provisioning, synchronization problems among the CR users etc for the CR users. The scheduling deals with an access protocol for which CR should use the proper channel at the proper time, if more than one CR users active in the network. In addition, if a PU is active suddenly in a network, definitely the CR should vacant the channel, so there should be a QoS provisioning for the cognitive terminals. As the radio spectrum is a very expensive resource and it requires efficient allocation to satisfy both licensed and unlicensed users, the economic issues involved in dynamic spectrum sharing are important.

## 4. SPECTRUM SENSING TECHNIQUES FOR CRs

The spectrum has been classified into three types as *black spaces*, *grey spaces* and *white spaces* by estimating the incoming RF stimuli (Wang, 2009). CR user may take the recompenses from grey spaces and white spaces of the spectrum. In order to hit upon those vacant spectrum spaces, the name of the a fundamental contestant is spectrum sensing. There are different methods proposed for identifying the presence of PUs signal transmission. The characterization of the transmitted signal enables not only to choose the signal transmission but also identifying the signal type. Generally, spectrum sensing techniques can be divided into three categories: *transmitter based detection* or *stand alone spectrum sensing, cooperative/distributed detection* and *interference based detection*.

### 4.1 Transmitter Based Detection

The cognitive user is liable to identify the presence of primary user in a certain geographic location and this is figure out in this type of detection. So, this is called transmitter based detection or stand alone detection (Bixio et al, 2009). This is a non-cooperative sensing technique which cannot detect the hidden primary user. This stand alone technique has been studied for military and civilian applications for signal detection, automatic modulation classification, to locate radio source and to perform the jamming activities in communication networks. In this section, some of the most common transmitter based sensing schemes in the CR literature are discussed. This scheme mainly includes *Energy detector based sensing*, *Matched filter*, *Waveform based sensing*, *Feature detection* and *Radio Identification based sensing*.

## 4.2 Energy detector based sensing

The fundamental scheme for spectrum sensing is based on energy detection (also known as Radiometry or Periodogram) where received signal energy is measured in a specific time epoch and certain frequency intervals. This technique is also prime choice for the engineers due to its low computational and implementation complexities. In addition, it is the most generic (as compared to methods given in this section) idea as receivers do not require any knowledge about the PUs signal (L. Bixio et al. 2009, p.185). The signal is detected by comparing the output of the energy detector with a threshold which depends on the noise floor. Some of the challenges with energy detector based sensing include selection of the threshold for detecting primary users, inability to differentiate interference from primary users and noise, and poor performance under low signal-to-noise ratio (SNR) values (Wang, 2009). Moreover, the dominant problem regarding energy detector as it does not work efficiently for detecting spread spectrum signals. However, this technique is the optimal solution when PUs transmission is not known (Wang, 2009). We want to formulate the energy detection based sensing as illustrated by Yücek and Arslan (2009). Let us consider that the received signal has the following simple form

$$y(n) = s(n) + w(n) \quad \ldots\ldots\ldots\ldots\ldots\ldots\ldots\ldots\ldots\ldots\ldots\ldots\ldots(1)$$

where *s(n)* is the signal to be detected, *w(n)* is the additive white Gaussian noise (AWGN) sample, and *n* is the sample index. If *s(n)* = 0, PU transmission is absent. The decision metric for the energy detector can be written as

$$M = \sum_{n=0}^{N} |y(n)|^2 \ldots\ldots\ldots\ldots\ldots\ldots\ldots\ldots\ldots\ldots\ldots\ldots\ldots(2)$$

where *N* is the size of the observation vector. The decision on the occupancy of a band can be obtained by comparing the decision metric *M* against a fixed threshold $\lambda_E$. This is equivalent to distinguishing between the following two hypotheses:

$$\mathcal{H}_0: y(n) = w(n) \ldots\ldots\ldots\ldots\ldots\ldots\ldots\ldots\ldots\ldots\ldots (3)$$
$$\mathcal{H}_1: y(n) = s(n) + w(n) \ldots\ldots\ldots\ldots\ldots\ldots\ldots\ldots\ldots (4)$$

The performance of the detection algorithm can be determined by two probabilities as the probability of detection, $P_D$ and probability of false alarm, $P_F$. $P_D$ is the probability of detecting a signal on the considered frequency when it truly is present. Thus, a large detection probability should be the desired consideration for the CR system. It can be formulated as

$$P_D = \Pr(M > \lambda_E | \mathcal{H}_1) \ldots\ldots\ldots\ldots\ldots\ldots\ldots\ldots\ldots\ldots (5)$$

Alternatively, $P_F$ is the probability that the test incorrectly decides that the considered frequency is occupied when it actually is not, and it can be written as

$$P_F = \Pr(M > \lambda_E | \mathcal{H}_0) \ldots\ldots\ldots\ldots\ldots\ldots\ldots\ldots\ldots\ldots (6)$$

$P_F$ should be kept as small as possible in order to prevent underutilization of transmission opportunities. The decision threshold $\lambda_E$ can be selected for finding an optimum balance between $P_D$ and $P_F$. However, this requires knowledge of noise and detected signal powers. The noise power can be estimated, but the signal power is difficult to estimate as it changes depending on ongoing transmission characteristics and the distance between the CR and PU (as cited in Yücek & Arslan, 2009). In practice, the threshold is chosen to obtain a certain false alarm rate. Hence, knowledge of noise variance is sufficient for selection of a threshold.

## 4.3 Matched filter

At a cognitive terminal, in order to maximize the output signal to noise ratio for a certain input signal a matched filter is designed which belongs to the linear filter (Wang, 2009, p. 410). Matched filter detection is applied if a cognitive user has *a-priori* knowledge of primary user transmitted signal. Therefore, Matched-filtering is known as the optimal strategy for detection of PUs in the presence of stationary Gaussian noise. The main advantage of matched filtering is the short time as it requires only O(1/SNR) samples to meet a given probability of detection constraint as compared to other methods that are discussed in this section. In fact, the required number of samples grows as *O*(1/*SNR*) for a target probability of false alarm at low SNRs for matched filtering. However, matched-filtering requires cognitive radio to demodulate received signals. Hence, it requires perfect knowledge of the primary users signaling features such as bandwidth, operating frequency, modulation type and order, pulse shaping, and frame format (Yucek & Arslan, p.123, 2009). Moreover, they discussed that CR needs receiver for all signal types, the implementation complexity of sensing unit is impractically large. In addition, this scheme consumes large power as various receiver algorithms require to be executed for detection. The problem to deal with Matched filter detection, requires a prior knowledge of PUs transmission criteria, if this information is not accurate, the matched filter performance degrades (Wang, 2009).

### 4.4 Feature detection

Another promising spectrum sensing technique is based on feature detection. An absolute information about the spectrum features can be obtained by identifying the transmission technologies used by primary users. Such an identification enables cognitive radio with a higher dimensional knowledge as well as providing higher accuracy (Yucek & Arslan, 2006). A feature captures a specific signature that is inherent characteristics of the primary user signal and the feature is unique for each class of signals. Some of the most common features considered are pilot signal, segment sync, field sync, and also the instantaneous amplitude, phase and frequency (Bixio et al, p. 186, 2009). This property is noticed in many of the signals employed in wireless communication and radar systems (W. Wang, p.411, 2009). There are several features that have been considered in order to detect and classify the primary user signals within a particular radio environment (as cited in Bixio et al, p. 186, 2009). Nowadays, analog to digital conversion has made the use of signal transform practical in order to detect a specific feature. Discrete Fourier transform, Wavelet transform and Wigner-Ville transform are belongs to this kind of ADC to detect the features of PU signal (Bixio et al, p. 186, 2009).

The fundamental and promising feature detection technique is based on the cyclic feature. This technique was introduced by Jondral et al (2007) and Sutton et al (2008) and they proposed the use of the cyclic feature as a reliable sensing technique for CR applications (Bixio et al, p. 186, 2009). Cyclic-feature detection approaches are based on the fact that modulated signal are usually coupled with sinusoidal carriers, hopping sequences, cyclic prefixes, spreading codes, or pulse trains, which result in a built-in periodicity (Bixio et al, 2009). These modulated signals are said to be cyclostationary. Cyclostationary features are originated by the periodicity in the signal in statistical manner like mean and autocorrelation (Gardner, 1991) or they can be intentionally used in order to sustain the spectrum sensing (as cited in Yucek & Arslan, 2009). This periodicity can be used as a feature and can be detected by analyzing a spectral correlation function (SCF), also known as cyclic spectrum (as cited in Bixio et al, 2009). Instead of power spectral density (PSD), cyclic correlation function is used for detecting signals present in a given spectrum. The cyclostationary based detection algorithms can differentiate noise from primary users' signals as the noise is wide-sense stationary (WSS) with no correlation while modulated signals are cyclostationary with spectral correlation due to the redundancy of signal periodicities. The cyclic spectral density (CSD) function of a received signal (1) can be calculated as suggested by (WA. Gardner, 1991).

$$S(f, \alpha) = \sum_{\tau=-\infty}^{\infty} R_y^{\alpha}(\tau) e^{-j2\pi f \tau} \ldots\ldots\ldots\ldots\ldots\ldots\ldots\ldots\ldots\ldots\ldots\ldots (7)$$

where

$$R_y^{\alpha}(\tau) = E[y(n+\tau)y^*(n-\tau)e^{j2\pi \alpha n}] \ldots\ldots\ldots\ldots\ldots\ldots\ldots\ldots\ldots (8)$$

is the cyclic autocorrelation function (CAF) and α is the cyclic frequency. The CSD function outputs peak values when the cyclic frequency is equal to the fundamental frequencies of transmitted signal x(n). Cyclic frequencies can be assumed to be known or they can be extracted and used as features for identifying transmitted signals (as cited in Yucek and Arslan, 2009). As a result, cyclostationary feature detector can overcome the energy detector limits in detecting signals in low SNR environments (Akyildiz et al, 2006). In fact, signals with overlapping features in the power spectrum, can have non-overlapping features in the cyclic spectrum (Gardner WA, 1988). Moreover, the cyclic spectrum is a much comfortable domain for signal detection than typical power spectral density. This property provides the flexible use of this technique as a more functional tool (Gardner WA, 1988).

Waveform based sensing is another promising feature detection scheme. Commonly used patterns like preambles, midambles, repeatedly transmitted pilot patterns, spreading sequences, etc, are utilized in wireless systems as a measure of synchronization or used for other purposes. A preamble is a known bit or signal pattern transmitted before each burst and a midamble is transmitted in the middle of a data frame. In the presence of a known pattern, sensing can be performed by correlating the received signal with a known copy of itself (as cited in Yucek & Arslan, p.122, 2009). This method is only applicable to systems with known signal patterns which is considered also a drawback of this kind of sensing, and it is termed as waveform-based sensing or coherent sensing. It is shown that waveform based sensing outperforms energy detector based sensing in reliability and convergence time. Furthermore, it is undoubtedly state that the performance of the sensing algorithm increases if the length of the known signal pattern increases. For analyzing the WLAN channel usage characteristics, packet preambles of IEEE 802.11b signals are exploited.

Another example of feature detection technique is while a PU is identified as a signal of any personal area networks (PANs) which belongs to Zigbee, Bluetooth, etc. CR can use this kind of information for extracting some useful information in space dimension as the range of signal of this type is in the very short range. Furthermore, CR may want to communicate with the identified communication systems in some applications. In the context of European transparent ubiquitous terminal (TRUST) project, radio identification, feature extraction and classification techniques are used (Farnham et al, 2000). The goal is to identify the presence of some known transmission technologies and achieve communication through them. As suggested by Farnham et al (2000), the two main tasks are initial mode identification (IMI) and alternative mode monitoring (AMM). In IMI, the cognitive device searches for a possible transmission mode following the power on whereas the later scheme monitoring the task of other modes while the cognitive device is communicating in a certain mode. In this sensing method, several features are extracted from the received signal and they are used for selecting the most probable PU technology by employing various classification methods. The features obtained by energy detector are used for signal classification. These features contains the energy distribution across the spectrum. Alternatively, channel bandwidth and its shape can be used as reference features. Channel bandwidth is found to be the most discriminating parameter among others. For classification, radial basis function (RBF) neural network is employed (as cited in Yucek & Arslan, 2009). In the research paper of Yucek and Arslan (2009), they illustrated the relevant complexity and accuracy for different types sensing scheme that is shown in the figure 4.

The OFDM waveform is altered before transmission in order to generate system specific signatures or cycle-frequencies at certain frequencies and effective signal categorization is being obtained with those signatures of the signal (as cited in Yucek & Arslan, 2009). Another research work done by Sutton et al (2007), investigated that in order to enhance the robustness against multipath fading, the number of features generated in the signal is increased. However, this comes at the expense of increased overhead and bandwidth loss. Hardware implementation of a cyclostationary feature detector is presented by Tkachenko et al (2007). The main advantage of the feature detection is that the discrimination between the noise energy and the modulated signal energy is very likely prepared. Moreover, cyclostationary feature detection can detect the signals with low SNR. In contrast, feature detection requires long observation time and higher computationally complex and also feature detection needs a-prior knowledge of the primary users (W. Wang, 2009).

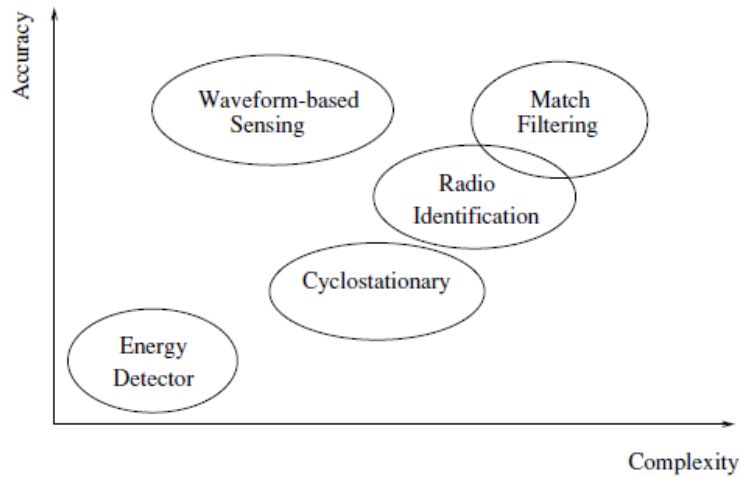

Fig. 4: Complexities and accuracies of different sensing methods (Yucek & Arslan, 2009)

### 4.5 Cooperative and distributed spectrum sensing

There are several problems like fading, shadowing and low SNR for desired performance associated to sense the signals through the stand alone transmitter and therefore cooperation is proposed as a solution to get rid of it. Cooperation is a sensing scheme which determine the existence of the PUs and the frequency band in use either and cooperation can be among CR users or external sensors can be used to build a common cooperative sensing network. The prime advantage of cooperation is it can solve the hidden terminal problem, when a CR is shadowed or in a deep fade (K. Arshad et al, 2010). In addition, the probabilities of misdetection and false alarm have considerably declined through the cooperative sensing of spectrum. In this sensing nature, several CRs can collaborate with each other in order to make a global decision about the existence of the PU to overcome this problem. Gandetto and Regazzoni (2007) proposed in their paper a distributed cooperative spectrum sensing by the involvement of a pair of cognitive terminals according to a distributed processing strategy. The proposed approach allow spectrum sensing by means of a "distributed cooperative consciousness" and the improved global performances of air interface detection is achieved than the single terminal spectrum approaches by using this method. In addition, it is already proved that collaborative spectrum sensing (CSS) can improve detection performance in the fading channels at the cost of increased computational complexity and bandwidth usage for exchanging information among CRs (as cited in Bixio et al, 2009).

In CSS, every cognitive user (CR) execute its own spectrum sensing tasks and make a local decision regarding the presence of a primary user. All of the CRs forward their decision to a common receiver, often called decision fusion centre or spectrum manager. For this reason, a dedicated feedback channel has to be allocated in order to share collected information which is termed as *reporting channel*. If the dedicated channel is absent, other methods should be considered which require less overhead (Bixio et al, 2009). The decision can be classified as soft (local measurement) or hard decision. In hard decision, fusion center collects binary decisions from the individual SUs, identifies the available spectrum, and then broadcasts this information to the other CRs. The optimal decision fusion is based on Neyman-Pearson criterion. In contrary, the soft decision, the value of sensing data transmission is higher to the fusion center through the reporting channels (K. Arshad et al, 2010). Fusion center may be an Access Point (AP) in wireless LAN or a CR base station in a cellular system and the fusion centre may be centralized or distributed. In centralized CSS, the fusion center assembles all the decision that comes from all the CRs and then make a decision that conveys to the CRs through the reporting channel. However, in distributed CSS, each of the CRs may behave as a fusion centre and receive sensing information from the neighboring nodes and take individual decisions.

Challenges of cooperative sensing include developing efficient information sharing algorithms and increased complexity (as cited in Yucek & Arslan, 2009). In cooperative sensing architectures, the control channel (pilot channel) can be implemented using different methodologies. These include a dedicated band, an unlicensed band such as ISM, and an underlay system such as ultra wide band (UWB). Depending on the system requirements, one of these methods can be selected. Control channel can be used for sharing spectrum sensing results among cognitive users as well as for sharing channel allocation information. As far as the networking is concerned, the coordination algorithm should have reduced protocol overhead and it should be robust to changes and failures in the network. Moreover, the coordination algorithm should introduce a minimum amount of delay.

Collaborative spectrum sensing is most efficient when the CR users observe fading or shadowing autonomously. The performance degradation due to correlated shadowing is investigated in terms of missing the opportunities. It is found that it is more advantageous to have the same amount of users collaborating over a large area than over a small area. In order to combat shadowing, beamforming and directional antennas can also be used. Cooperative sensing network can be constructed by the presence of cognitive radios or external sensors may perform sensing tasks together. In the former case, cooperation can be implemented in two fashions: centralized or distributed (as cited in Yucek & Arslan, 2009 ).

### 4.5.1 Centralized Sensing

Let a set of *N* cognitive user share the same radio channel. Each CR perform the sensing individually by any of the technique discussed in the earlier section (Bixio et al, 2009). In centralized sensing, a central unit collects those sensing information from cognitive devices, identifies the available spectrum, and broadcasts this information to all the cognitive radios and also the decision of spectrum in use by the CRs. This means it can directly controls the cognitive radio traffic. The sensing results (hard or soft decision) are gathered at a central place which is known as access point (AP) or a cognitive manager. The goal is to mitigate the fading effects of the channel and increase detection performance. Hard and soft information combining methods are investigated for reducing the probability of missed opportunity. In some cases, users send a quantized version of their local decisions to central unit (fusion center). Then, a likelihood ratio test (e.g., Neyman-Pearson test) over the received local likelihood ratios is applied. In contrast, for large number of CR users, the traffic as well as bandwidth required for reporting channel becomes huge. Therefore, local observations of cognitive radios are quantized to one bit (hard/binary decision) to reduce the sharing bandwidth. There are two advantages associated with the hard decision; first it gives relief from the traffic to the reporting channel and later is the energy efficient sensing as for the quantization and transmission of one bit, required energy is lesser than usual. Furthermore, only the cognitive radios with reliable information are allowed to report their decisions to the central unit (Yucek and Arslan, 2009).

### 4.5.2 Distributed Sensing

In the case of distributed sensing, cognitive nodes share information among each other but they make their own decisions as to which part of the spectrum they can use. Distributed sensing is more advantageous than centralized sensing in the motive that there is no need for a backbone infrastructure and it has reduced cost. It is necessary to include in this context, different solutions (distributed detection with or without fusion) can be used depending on the level of cooperation among the CRs. Distributed detection with data sensing scheme provide better performance than the stand alone transmitter based detection. Several problems associated with the distributed detection but centralized data fusion for the CR applications:

- a dedicated reporting channel is needed to share the information which may not be available, if available, the reporting channel can be faded or shadowed.
- high computational capabilities required at the data fusion center.

In order to deal with the above listed issues, distributed detection without fusion spectrum sensing techniques have been developed. As this technique is based on "implicit" cooperation among terminals and it does not require any dedicated channel (Bixio et al, 2009). In these approaches a set of *N* cooperative cognitive radios share the same radio environment. Each cognitive radio performs spectrum sensing based on its local observation. These local decisions are not fused to obtain a global decision and no sharing of information is required. Yucek and Arslan (2009) introduced an incremental gossiping approach termed as GUESS (gossiping updates for efficient spectrum sensing) in order to perform efficient coordination between CRs in distributed collaborative sensing. They proposed in their algorithm with minimizing complexity as well as reduced protocol overhead.

Sometimes, collaboration for information exchanging is between two CR users. The CR user close to a primary transmitter, which has a better chance of detecting the primary user transmission, cooperates with far away users. An algorithm for pairing secondary users without a centralized mechanism is proposed. A distributed sensing method where CR users share their sensing results among themselves and here only final decisions are shared in order to minimize the network overhead due to collaboration (as cited in Yucek & Arslan, 2009). Features obtained at different radios are shared among cognitive users to improve the detection capability of the system.

### 4.6 External Sensing

Another technique for obtaining spectrum information is external sensing. In external sensing, an external agent (sensor) performs the sensing and broadcasts the channel occupancy information to cognitive radios. External sensing algorithms solve some problems associated with the internal sensing where sensing is performed by the cognitive transceivers internally. Internal sensing is termed as collocated sensing. The main advantages of external sensing are overcoming hidden primary user problem and the uncertainty due to shadowing and fading. Furthermore, as the cognitive radios do not spend time for sensing, spectrum efficiency is increased. The sensing network does not need to be mobile and not necessarily powered by batteries. Hence, the power consumption problem of internal sensing can also be addressed. The presence of passive receivers, *viz.* television receivers, is detected by measuring the local oscillator power leakage. Once a receiver and the used channel are detected, sensor node notifies cognitive radios in the region of passive primary users via a control channel. A dedicated network composed of only spectrum sensing units is used to sense the spectrum continuously or periodically. The results are communicated to a sink (central) node which further processes the sensing data and shares the information about spectrum occupancy in the sensed area with opportunistic radios. These opportunistic radios use the information obtained from the sensing network for selecting the bands (and time durations) for their data transmission. A pilot channel can be used to distribute the sensing results which is similar to network access and connectivity channel (NACCH). External sensing is one of the methods proposed for identifying primary users in IEEE 802.22 standard as well.

To this end, there is another sensing method named *interference based sensing* available in the literature. This interference temperature model was introduced by FCC and this can be defined as "*the temperature equivalent of the total interference present in RF environment for a particular frequency band and a certain geographic location*" (W. Wang, 2009). The CR measure the interference temperature and regulate their transmission in such a way that they avoid raising the interference temperature over the interference temperature limit. Hidden terminal problem can be avoided by implementing this scheme. Instead, difficulties for the measurement of interference temperature put this scheme into a challenge and during detection the CRs cannot differentiate the signals coming from the PU or noise

## 4.8 Energy efficient spectrum sensing techniques for the green CR network

Nowadays, existing research is going to address the distributed spectrum sensing (DSS) scheme and the challenges of it. In order to perform decision combining, the CR users collect the information from their neighbors and make the decision autonomously through the reporting channel. However, if the *reporting channel* experiences fading, the sensing performance degrades significantly (Wei & Zhang, 2010). The activated CR users transmit the spectrum sensing results by broadcasting, which is energy-expensive if the secondary users are spread out in a wide area. In 2007, Letaief et al introduced a cluster based DSS method by selecting the secondary user with the largest reporting channel gain as the cluster head (CH) of each cluster. Though they showed the traditional clustering methods which is not sufficiently efficient in terms of the energy consumption. Wei and Zhang (2010) they proposed a cluster and forward based DSS scheme which is composed by two-tier hierarchical CR network, is discussed in the following sections.

### 4.8.1 Cluster based data fusion

In that scheme, all the secondary users exist in a clustered based geographic location. A pre-defined threshold is chosen to gather the CR users sensing contribution results and the CR user send its spectrum sensing result when its contribution is positive. After each sensing round, all the CR users know the majority sensing decision made by their neighbors. The CH is elected with the largest contribution value among the CR users, which collects and processes the local information sent from its cluster members. The clustering and CH selection at each time step, and the active CHs take turns to be the fusion center is proposed in order to make the CR system energy efficient (Wei & Zhang, 2010).

### 4.8.2 Contribution based decision scheme

Consider that each available CR user gets the decisions from the others. Because of the energy constraint, the CR users exchange the simple message than transferring the complicated information, as proposed by Wei and Zhang, (2010). In this scheme, each CR user cannot send its spectrum sensing result until it provide the positive contribution, and in this way each CR user knows the majority decision of the spectrum sensing at the end of each spectrum sensing round. If the CR user obtains the spectrum sensing result consistent with the consensus, it has the positive contribution; otherwise negative contribution can be considered. A pre-defined threshold for detection is chosen in the scheme. When the CR user's contribution value drops below the threshold, this CR user is considered unreliable and stops it from sending its spectrum sensing results. However, this CR still carrying spectrum sensing and tracking the majority decision. As long as its own spectrum sensing result agrees with the majority decision, its contribution is evaluated. As long as the contribution value exceeds the pre-defined threshold, it starts sending the spectrum sensing results again.

### 4.8.3 Cluster and forward scheme

To make the CR network reliable, DSS should send its spectrum sensing decision to its CH. As broadcasting the decisions be an expensive method in the context of energy if the CR users are spread out in a wide area, the forwarding method can be chosen as it requires less energy according to the inverse square-law of power transmission. In the cluster-and-forward scheme, all active CHs and cluster members judged trustworthy as proposed by Wei and Zhang, (2010). Firstly, clustering is done based on geometric locations and the CR users who are close together form a cluster. Then, one of the cluster members with the highest spectrum sensing ability is selected as the cluster head (CH), which processes the spectrum sensing results sent from its cluster members and also collects the individual spectrum sensing decisions. Moreover, due to the energy constraint, the CH is re-selected at each time step as to avoid draining a secondary user's power quickly. After acquiring the spectrum sensing results by the cluster members and also to obtain the spectrum sensing decisions, they transmit not only the results to the selected CH but also the CH adds its own sensing decision and then forwards the all the results of the sensing to the fusion center. In the proposed scheme, the fusion center is dynamically selected from all the active CHs and hence energy saving is achieved by clustering as well as combining data before forwarding. There are some other considerations about the clusters chosen for the sensing scheme, i.e. if the cluster range is small enough so that the local spectrum sensing decision and the local contribution-value collection consume less power while shorter range of clustering increase number of clusters. The more the clusters make the system running in parallel and less delay. However, the more clusters impacts on the data overhead, which are undesirable.

### 4.8.4 Compressive sensing techniques

Narrow-band spectrum sensing is simpler to implement than the wide-band applications which are nowadays increasingly interested as the deployment of dynamic spectrum access for CR users (Tan & Kong, 2010). However, wide-band spectrum sensing introduces with considerable technical challenges. This wide band sensing can be done in two different ways as the RF front-end can either do narrow-band sensing via a bank of pass-band filters or implement one wide-band RF front-end followed by signal processing blocks to sense the whole wide-band. Nonetheless, the previous associated with the filter design constraints while the later scheme requires high-speed analog-to-digital converter. Recently, potential approach to alleviate the sampling bottleneck in wideband communications is *compressive sensing* (CS), which asserts that the one can recover sparse signals at sub-Nyquist rates (as cited in S. Hong, 2010). CS relies on this principle of sparsity, so that a concise representation of the signal is possible when expressed in a suitable from. Due to the low percentage of spectrum occupancy by PUs, which originally motivated the development of CRs, wireless communication signals in open-spectrum networks are typically sparse in the frequency domain, allowing us to use compressive sensing to alleviate the sampling bottleneck. While CS is a powerful technique, it does not allow the CR to sample at low rates for free; the resulting increase in computation/complexity is non-trivial, especially in a power constrained mobile cognitive radio. Recent work in CS reveals that a signal having a sparse representation in one basis can be recovered from a small number of projections. Especially, a compressed sampling approach can get the sparse signal at the rates lower than Nyquist sampling; signal reconstruction, which is a solution to a convex optimization problem (Tan & Kong, 2010). As the sampling is done less than the Nyquist rate, so the signals requires less energy for processing at both the transmitter and receiver ends which is supportive to move forward for green cognitive radio.

### 5 POWER CONTROL IN COGNITIVE TRANSCEIVER

The model of DSA techniques, although overall spectrum utilization can be enhanced through exploiting it while radio signal transmission from cognitive nodes can cause harmful interference to the PUs which validates two important design considerations for CR networks, i.e. to maximize the radio resource utilization and to minimize the interference caused to the PUs (Hoang & Liang, 2008). Therefore, power control in CR communication should be necessary regarding the probable interference originated since the presence of multiple users, either primary or secondary, within this network. Generally, power control is employed in mobile networks for improving the link performance that can be applied to CR networks as well, however, power control in CR network is a crucial task and this is the basic module of radio resource management (Qin & Su, 2009). In a wireless network where users are prioritized (i.e. both PUs and CRs are present), if a CR gets the access to exploit the spectrum, it must taking into account the power transmission itself to assure the interference-free transmission for users (PU and CR both) surrounding in it. There are different techniques of this kind, for example, spectrum underlay is one of the existing techniques which permit the CR users to share the whole spectrum simultaneously with the PUs as long as they do not exceed the interference threshold at PUs (Pareek et al, 2010). An excellent survey on power control of CR has been demonstrated and this is followed the so called "*protocol model*" for interference modeling (Shi & Hou, 2007). Since power control directly affects a node's transmission range and interference range (receive power at the destination node and interference power at other nodes), it has profound impact on scheduling feasibility, bandwidth efficiency, and problem complexity (Shi & Hou, 2008). Furthermore, power control is an important issue for the energy-efficient spectrum sensing, the future trend of green cognitive radio networks.

### 5.1 Power control for spectrum sharing

It can be said that implementation of DSA technique can improve the overall spectrum efficiency though transmission from cognitive devices can cause harmful interference to primary users of the spectrum (Hoang & Liang, 2008). The transmitter power plays an important role in the radio resource management in CR networks as it limits multiuser interference and to maximize the spatial resource reuse (Qin & Su, 2009). Clearly, for the network layer, power control impact the network topology. For the MAC layer, power control also affects how far apart can two ongoing communication sessions be without interfering with each other while at the physical layer, power control is linked to the signal processing, because the signal processing at the physical layer identifies how stringent the power control requirements to be. All these factors determine the end-to-end performance (Qin & Su, 2009). As a result, there are many cross-layer design methods in the literature have looked at power control in a cross-layer design framework. With the increasing demand for wireless data services, it is necessary to establish power control method for information sources. The power control solution for wireless data solved by the game theoretic approach as a pricing function of the transmit power. Game theory can be so applied to power control in CDMA wireless networks as discussed in (Saraydar et al, 2002).

### 5.2 Spectrum sharing using decentralized scheduling

As proposed by Qin and Su (2009) considering $N$ primary services going on within a CR network of different frequency spectrum $F_i$ and a secondary service is responsible to give service to a set of CR users to share the spectrum with some primary services. The spectrum can be shared among multiple secondary users, where the base station or access point manages the allocated spectrum's radio transmission. In the allocated frequency bands, the CR users' spectrum demand relies on the transmission rate for the adaptive modulation.

In the decentralized scheduling, each node has two spectrum management decisions: a *diverse transmission power control* with local *decentralized scheduling* and a *decision maker* with the acceptance and distribution of frequency bands among participating CRN nodes. As proposed by Qin and Su, (2009) frequency spectrums are submitted by a local user community through the node's decentralized scheduling and flow routing. With the *decision maker*, it is queued and either started locally or moved to other nodes. The onsite allocation to primary services is eventually done by the local diverse transmission power control. Firstly, the decision maker accepts spectrum from a local user community, it then either allocates frequency bands to the local node's decentralized scheduling and flow routing or forwards them to the decision maker of other nodes.

### 5.3 Design of distributed optimization algorithm

In a wireless network, each node is associated with a certain level of interference range that it can tolerate, which occurs with two pairs of nodes of transmissions that overlap in frequency and time. When a node observes the interference as a low signal to interference and noise ratio (SINR) then it should increase the transmission power to enhance the SINR and through this way each radio should maximize SINR at the anticipated receiver. Undoubtedly, this strategy does not suit well as increasing the transmitter power from each radio than a predefined level results increasing power consumption, decreasing battery life, and raising potential interference to others. To overcome this problem, Qin and Su, (2009) proposed the techniques ***scheduling-power strategy*** and ***global solution with diverse transmission power*** which select the SINR at a suitable level while maintaining the interference under control. In the earlier strategy, the two-phrase power control with decentralized scheduling (PCDS) algorithm is proposed. At the beginning of each time slot this algorithm checks the interference level developed in some slots. Without interrupting other transmission, the PCDS determines the users set which can transmit in the current slot in security. Basically, it performs two tasks: one is to identify the users set which can try to transmit in a given time slot, the other is to determine the powers set needed to satisfy SINR constraints at respective receivers. The supplementary strategy deals with the process of frequency bands exchange on two different scenarios: the active delegation asks the local decision maker to provide frequency bands to remote nodes in an active way, and the passive delegation asks remote nodes to request frequency bands for the local decision maker. In each scenario, the local decision maker has to publish information of its local waiting queue. The global solution with diverse transmission power builds a connected network, but it does not set all transmission ranges to the same value. Instead, for every node, it tries to find a minimum power level individually.

### 5.4 Power control in multiuser CR networks

Recently, joint beamforming and power control techniques are widely considered for multiuser CR systems. Whitening transform is used for simplifying the capacity optimization, and classic water-filling method is used for beamforming and power control. Wang and Wang (2010) discussed into their paper for the multiuser CR network, they addressed the problem of joint transmit beamforming and power control for CR users when they are allowed to transmit simultaneously with the primary users. By using joint transmit beamforming, sum rate optimization of the network is achieved under the interference constraints of PUs. A transmit beamforming technique is discussed to obtain the achievable rates for the multi-antenna cognitive radio network (Bixio et al, 2011).

## 6   CONCLUDING REMARKS AND POSSIBLE FUTURE SCENARIOS

This chapter overviewed the state-of-the-art techniques for dynamic spectrum management and spectrum sensing for the cognitive radio network as they were proposed so far in literature. The main purpose of this work is to underline the pros and cons of various described techniques. The various proposed schemes have been reviewed and compared having in mind the real-life applications. Dealing with green CR network, several energy-efficient sensing schemes have been proposed. In addition this chapter pointed out how the transmitter power plays an important role in the radio resource management for CR networks; power control optimization was shown as it is needed for limiting multiuser interference as well as for maximizing spatial resource reuse and desired throughput.

Research is still carried out for deploying the dynamic spectrum management: spectrum sensing with single node detection is survived with several problems which could be solved by distributed sensing methods while the computational complexity and hardware constraints push those schemes into challenge. When considering centralized and distributed sensing, the optimization technique should be implemented for both data and decision fusion. There is a possibility for the future wireless network:- flexible radio, which will be increasingly complex and certainly heterogeneous in nature. The concept of flexible radio plays vital role in the future wireless communication (3G-long term evolution: LTE, and LTE-advanced) that must satisfy the adaptability, reconfigurability, modularity, scalability and so on.